\documentclass[journal]{IEEEtran}

\usepackage{cite}
\usepackage[dvips]{graphicx}
\usepackage[cmex10]{amsmath}
\usepackage{array}

\usepackage{mdwmath}
\usepackage{mdwtab}

\usepackage{eqparbox}

\begin{document}

~
\vspace{5cm}

\copyright[2009]IEEE.  Reprinted, with permission, from:

{\it Wideband Reference-Plane Invariant Method for Measuring
Electromagnetic Parameters of Materials}

Authors: Chalapat, K.; Sarvala, K.; Jian Li; Paraoanu, G.S.;

Journal: Microwave Theory and Techniques, IEEE Transactions on

Volume: 57 , Issue: 9

Digital Object Identifier: 10.1109/TMTT.2009.2027160

Publication Year: 2009 , Page(s): 2257 - 2267

\vspace{2mm}
This material is posted here with permission of the IEEE.  Internal or
personal use of this material is permitted.  However, permission to
reprint/republish this material for advertising or promotional purposes or
for creating new collective works for resale or redistribution must be
obtained from the IEEE by writing to pubs-permissions@ieee.org.

By choosing to view this document, you agree to all provisions of the
copyright laws protecting it.

\newpage

\title{Wideband Reference-Plane Invariant Method for Measuring Electromagnetic Parameters of Materials}

%
%
%

\author{Khattiya~Chalapat,
        Kari~Sarvala, Jian~Li and~Gheorghe~Sorin~Paraoanu
		
\thanks{Manuscript received June 25, 2008; revised March 19, 2009. This work was supported by Thailand Commission on Higher Education and the Academy of Finland (Acad. Res. Fellowship 00857 and projects 7111994, 7118122 and 129896).}
\thanks{K. Chalapat and K. Sarvala were with Nanoscience Center, Department of Physics, University of Jyv\"{a}skyl\"{a}, Finland. K. Chalapat is now with the Low Temperature Laboratory, Helsinki University of Technology (e-mail: khattiya.chalapat@ltl.tkk.fi, KChalapat@gmail.com).}
\thanks{K.~Sarvala is also with Nokia Oyj, FI-40100 Finland.}
\thanks{J.~Li and G.~S.~ Paraoanu are with Low Temperature Laboratory, Helsinki University of Technology, P.O. Box 5100, 
FI-02015 Finland.}}


%
%

\markboth{}%
{Chalapat \MakeLowercase{\textit{et al.}}: }
%



\maketitle

\begin{abstract}
 This paper presents a simple and effective wideband method for the determination of material properties, such as the
 complex index of refraction and the complex permittivity and permeability. The method is explicit (non-iterative) and reference-plane invariant: it uses a certain combination of scattering parameters in conjunction with
 group-velocity data. This technique can be used to characterize both dielectric and magnetic materials.
 The proposed method is verified experimentally within a frequency range between 2 to 18 GHz on polytetrafluoroethylene and polyvinylchloride samples. A comprehensive error and stability analysis reveals that, similar to other methods based on transmission/reflection measurement, the uncertainties are larger at low frequencies and at the Fabry-P\'erot resonances.

\end{abstract}


%
\IEEEpeerreviewmaketitle

\section{Introduction}
\IEEEPARstart{I}{n} recent times, fast and accurate knowledge of the electromagnetic properties of materials in the microwave range is increasingly required in the design and development process of a vast number of industries, spanning from food processing to communication systems. Also in basic science research, understanding and measuring material parameters, such as the complex refractive index $n$ and the complex permittivity $\epsilon$, is an important fundamental task.

Nonresonant techniques such as transmission and reflection measurements are largely used nowadays for characterizing the electromagnetic properties of materials \cite{book}; the fundamentals of these techniques have been already established in the 1970's by the seminal papers of Nicolson, Ross \cite{NicolsonRoss} and Weir \cite{Weir}. These techniques are relatively simple and accurate. They have the advantage of broadband characterization of materials and devices. For almost four decades they have been widely applied to measure the permittivity and permeability of various synthetic and natural materials.

A known drawback of the original Nicolson-Ross-Weir (NRW) method is that it requires the transformation of S-parameter measurements from the calibration reference planes to the surfaces of the material. The phases of the transmission and reflection signals are strongly dependent on the positions of the reference planes, so the uncertainties in the transformation of S-parameters can add significant errors. The precision of this transformation can be enhanced in various ways, for examples by adding more steps to the calibration process or running extra calculation algorithms which complicate the measurement. Besides the Nicolson-Ross-Weir algorithm, other methods based on transmission/reflection measurements exhibit the same sort of difficulties.

An important step forward has been achieved in 1990, when Baker-Jarvis and collaborators showed that it is possible to derive S-parameter equations which are reference-plane invariant \cite{BakerJarvis}. Using some of these equations, they showed that it is possible to extract the values of the material parameters by using an iterative algorithm. This algorithm requires as input some initial values for permittivity and permeability, and therein lies one of its limitations: a good guess is needed, otherwise the algorithm can produce wrong results. Although valuable work has been done recently to improve these ideas \cite{StuchlyMatuszewski}-\cite{Hasar}, there is currently no universally-accepted best method. Each proposed technique has a number of advantages and disadvantages and the weight to be attached to each depends on the specific application.

The purpose of this paper is to present a methodology for extracting the complex material parameters from transmission/reflection measurements which combines ideas from the NRW and the Baker-Jarvis techniques. More precisely, the scattering parameters are combined into a specific set of reference-plane invariant equations (similar to Baker-Jarvis), and the equations are used together with group velocity data (similar to NRW) to obtain the complex permittivity and permeability. Surprisingly, this results in a simple, explicit, and reference-plane invariant methodology which can be used to characterize both dielectric and magnetic materials. With respect to NRW, the advantage of our method is that it uses reference-plane invariant quantities, therefore the errors due to calibration to the two material-air interface are eliminated. With respect to the Baker-Jarvis algorithm, the improvement consists in the use of additional information about the sample, extracted from group velocity measurements. This removes the ambiguity in the determination of the phase. Also, since our results do not depend on choosing good initial values for $\epsilon$ and $\mu$ as in Baker-Jarvis\footnote{The method of Baker-Jarvis is tailored for dielectric materials, thus one takes $\mu_{r}=1$ as initial value.}, dielectric materials with unknown properties and also materials with magnetic properties at high frequency can be characterized.

\section{Theory}

In this section, the measurement is modeled within the framework of classical electrodynamics. We present a new algorithm which is reference-plane invariant and show how it can be used to determine the complex refractive index and the complex permittivity and permeability.

\begin{figure}[!h]
\centering
\includegraphics[width=2.5in]{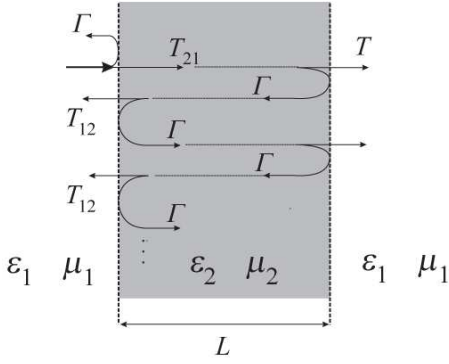}
\caption{The model of multiple reflection between two interfaces.}
\label{MultipleReflection}
\end{figure}

We start by deriving the mathematical relations between
the S-parameters and the material parameters.
We describe the scattering of electromagnetic waves based on the multiple
reflection model shown graphically in Fig. \ref{MultipleReflection}.
Within this model, the total reflection and transmission coefficients can be
calculated using the superposition
principle. Since a transverse electromagnetic (TEM) wave traveling a
distance $L$ picks up a phase change of $2 \pi
L/\lambda$, where $\lambda$ is the wavelength in that region, the
propagation factor of the TEM wave traveling through
the material of length $L$, as shown in Fig. \ref{MultipleReflection} is given by
\begin{equation}
P = e^{-\gamma_2 L},
\label{Psec2}
\end{equation}
where $\gamma_2 = i\omega /v_2 = i\omega n_2/c = i\omega \sqrt{\mu_{2}\epsilon_{2}}$.

The total reflection coefficient is
\begin{eqnarray}
\Gamma_{\rm tot} &=& \Gamma + T_{21}T_{12}\Gamma P^2 + T_{21}T_{12}\Gamma^3 P^4 + ... \nonumber \\
&&\\
&=&\frac{\Gamma (1-P^2)}{1- \Gamma^2 P^2},\nonumber
\label{Gamma_tot}
\end{eqnarray}
where
\begin{equation}
\Gamma = \frac{1-\sqrt{\frac{\epsilon_2 \mu_1}{\epsilon_1 \mu_2} }}{1+\sqrt{\frac{\epsilon_2 \mu_1}{\epsilon_1 \mu_2} 
}},
\label{GammaDef}
\end{equation}
and
\begin{equation}
T_{12} = 1 + \Gamma = \frac{2}{1+\sqrt{\frac{\epsilon_1 \mu_2}{\epsilon_2 \mu_1} }} = \sqrt{\frac{\epsilon_2 
\mu_1}{\epsilon_1 \mu_2}} \hspace{2pt}T_{21}. \label{T12}
\end{equation}

Similarly, the total transmission coefficient in terms of $\Gamma$ and $P$ is
\begin{equation}
T_{\rm
tot} = \frac{P \left(1 - \Gamma^2\right)}{1-\Gamma^2 P^2}.
\label{Teq4}
\end{equation}

\begin{figure}[!h]
\centering
\includegraphics[width=2.5in]{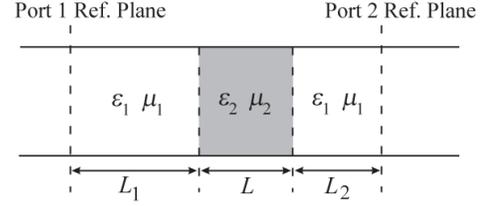}
\caption{The model of a transmission line containing a material of length $L$. $L_j$ ($j=1,2$) represents the distance from the 
reference plane of the S-parameter measurement to the corresponding interface between air and the material under test.}
\label{TransmissionReflectionMethod}
\end{figure}

The standard model of a TR measurement is described
in Fig. \ref{TransmissionReflectionMethod}. The transmission line is divided into three regions. Typically, the regions 
of lengths $L_1$ and $L_2$ are assumed to be filled with air and the middle region of length $L$ is filled with a 
material of relative permittivity $\epsilon_r = \epsilon/\epsilon_0$  and relative permeability $\mu_r = 
\mu/\mu_0$. The complex refractive index of the material is $n = \sqrt{\mu_r\epsilon_r}$.
We then take the permittivity, $\epsilon_1$, and the permeability, $\mu_1$, to be equal to the permittivity of free space, $\epsilon_0$, and, respectively, the permeability of free space, $\mu_0$, and write Eq. (\ref{GammaDef}) in the form

\begin{equation}
\Gamma = \frac{z-1}{z+1},
\label{Gamma}
\end{equation}

where $z = \sqrt{\mu_{r}/\epsilon_{r}}$ is the impedance relative to vacuum (the total impedance
$Z=\sqrt{\mu/\epsilon} = Z_{0}z$, where $Z_{0}= \sqrt{\mu_{0}/\epsilon_{0}}=120\pi\Omega$ is the vacuum impedance).

The determination of these two quantities, $z$ and $n$, from the experimental data will be the main focus of the remaining part  of this section. Based on the multiple reflection model, the S-parameters are expressed in terms of $\Gamma$ and $P$ as follows
\begin{equation}
S_{11} = e^{-2\gamma_1 L_1}\Gamma_{\rm tot} = e^{-2\gamma_1 L_1} \frac{\Gamma (1-P^2)}{1-\Gamma^2 P^2},
\label{S11sec2}
\end{equation}
\begin{equation}
S_{22} = e^{-2\gamma_1 L_2} \Gamma_{\rm tot} = e^{-2
\gamma_1 L_2} \frac{\Gamma(1-P^2)}{1-\Gamma^2 P^2},
\label{S22sec2}
\end{equation}
and
\begin{equation}
S_{21} = S_{12} = e^{-\gamma_1 (L_1 + L_2)}T_{\rm tot} = e^{-\gamma_1 (L_1 + L_2)} \frac{P (1-\Gamma^2)}{1-\Gamma^2 P^2}.
\label{S21sec2}
\end{equation}
where $\gamma_1 = i\omega n_1/c \approx i\omega /c$ ($c$ is the speed of light in the vacuum).

When there is no sample inside the transmission line, $\Gamma=0$, $\gamma_2 = \gamma_1$ and therefore
\begin{equation}
S_{21}^o = e^{-\gamma_1 (L_1 + L_2 + L)}=e^{-\gamma_1 (L_{\rm air})}.
\label{S21o}
\end{equation}
Eq. (\ref{S21o}) allows us to experimentally determine the airline length, $L_{\rm air}$, by calibrating the vector network analyzer and then measuring the transmission through the empty air line to obtain the phase of $S_{21}^o$.

In the next step, $\Gamma$ and $P$ are expressed in terms of the S-parameters. This is similar to the Nicolson-Ross-Weir
algorithm, with the essential difference that neither $\Gamma$ nor $P$ depends on $L_1$ and $L_2$. Indeed, for 
airlines operating at relatively high frequencies, measurements of $L_1$ and $L_2$ are prone to relatively large 
uncertainties. These errors will further propagate in the phase factors of the S-parameters,
\begin{equation}
\delta \left(2\gamma_{1}L_j\right) =  \frac{i 4\pi f  \delta L_j}{c}, {\rm \hspace{0.5cm}} j\in \{1,2\},
\end{equation}
leading to the larger errors of the phase factors at higher frequencies.

\begin{figure}[!h]
\centering
\includegraphics[width=2.5in]{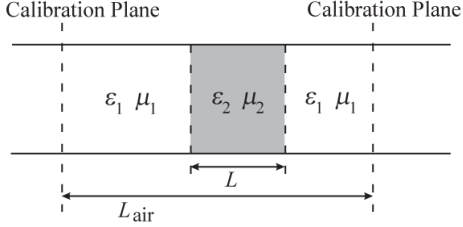}
\caption{Reference-plane invariant measurement model. In the diagram, $L$ represents the length of the sample and 
$L_{\rm air}$ represents the distance between the calibration planes of the S-parameter measurement.}
\label{RefPlaneInvariantMethod}
\end{figure}

Fig. \ref{RefPlaneInvariantMethod} shows the schematic model of a reference-plane invariant measurement. We start by defining two quantities, which are related to measurable quantities, namely
\begin{equation}
A = \frac{S_{11}S_{22}}{S_{21}S_{12}} = \frac{\Gamma^2}{(1-\Gamma^2)^2}\frac{(1-P^2)^2}{P^2},
\label{Adef}
\end{equation}
and
\begin{equation}
B = e^{2 \gamma_1 (L_{\rm air} - L)}(S_{21}S_{12}-S_{11}S_{22}) = \frac{P^2 - \Gamma^2}{1- \Gamma^2 P^2}
\label{Bdef} .
\end{equation}

Experimentally the $S$-parameters are measured by a vector network analyzer (VNA), the airline length, $L_{\rm air}$, is
found via Eq. (\ref{S21o}), and the length of the sample, $L$, is measured before inserting the sample into the airline.

Solving Eq. (\ref{Bdef}) for $P^2$,
\begin{equation}
P^2 = \frac{B +  \Gamma^2}{1 + B \Gamma^2},
\label{P2BC}
\end{equation}

\noindent and substituting back into (\ref{Adef}), we obtain

\begin{equation}
A = \frac{\Gamma^2 (1 - B)^2}{(B + \Gamma^2)(1 + B \Gamma^2)}.
\label{ABCGamma}
\end{equation}
Eq. (\ref{ABCGamma}) can be solved to find $\Gamma^2$
\begin{eqnarray}
\Gamma^2&=&\frac{-A(1+B^2)+(1-B)^2}{2AB}\nonumber\\ &&\label{Gamma2ABC}\\
&&\pm \frac{\sqrt{-4A^2B^2+\left[A(1+B^2)
-(1-B)^2\right]^2}}{2AB}, \nonumber \\ \nonumber
\end{eqnarray}
where the sign in this equation is chosen so that  $|\Gamma| \leq 1$. These expressions for $P^2$ and $\Gamma^2$ are manifestly
reference-plane invariant.

A very useful, simpler expression for $P$ can be obtained if we define another quantity, $R$, directly related to the scattering parameters,
\begin{equation}
R = \frac{S_{21}}{S_{21}^o} = \frac{e^{\gamma_1 L} P (1-\Gamma^2)}{1-P^2\Gamma^2}.
\label{Ddef}
\end{equation}
Then Eq. (\ref{Ddef}) can be solved for $P$. Substituting $P^2$ from (\ref{P2BC}) into the denominator of (\ref{Ddef}), we
obtain

\begin{equation}
P = R\frac{1+\Gamma^2}{1+B\Gamma^2}e^{-\gamma_{1}L}
\label{PRGamma}
\end{equation}

\noindent From (\ref{Psec2}), assuming free space on either side of the sample, the complex refractive index can be determined by

\begin{equation}
n = \sqrt{\epsilon_r\mu_r} = \frac{1}{\gamma_1 L} \ln\left(\frac{1}{P}\right),
\label{n}
\end{equation}
or, more explicit,
\begin{equation}
n = 1 -\frac{1}{\gamma_{1}L}\ln\left(\frac{1+\Gamma^2}{1+B\Gamma^2}R\right).
\label{nRefPlaneInvariant}
\end{equation}

The logarithmic function in (\ref{n}) and (\ref{nRefPlaneInvariant}) is a multi-valued function, which results in an infinite number of discrete values for $n$ . The physically correct solution must be chosen from these values. One way to do so is to check whether a chosen solution gives correct values for another measurable quantity or not. Also, this measurable quantity should not depend on  $L_1$ and $L_2$. In the following, we will show that an appropriate such quantity is the group delay in the line.

By definition, the group delay is a measure of a pulse signal transit time through a transmission line. The transit time of a wave packet is defined as
\begin{equation}
\tau_{\rm{g}} = \frac{x}{v_g},
\label{Tg_def}
\end{equation}
where $x$ is the transit length and $v_g$ is the group velocity of the wave pulse. In this case,
\begin{equation}
\tau_{\rm{g}}^o = \frac{L_{\rm air}}{c},
\label{Tg_empty}
\end{equation}
and
\begin{equation}
\tau_{\rm{g}} = \frac{L_{\rm air}-L}{c} + L \frac{d}{df}\left(\frac{f n}{c}\right).
\label{Tg_cal}
\end{equation}
where $\tau_{\rm{g}}^o$ is the group delay through an empty line, and $\tau_{\rm{g}}$ is the group delay through the
line with an inserted sample of length $L$. By comparing the calculated group delay with the measured group delay,
\begin{equation}
\left| \tau_{\rm g}^{\rm (measured)} - \tau_{\rm g}^{\rm (calculated)} \right| = 0,
\end{equation}
the correct refractive index can be determined. Another quantity which can be used as an alternative to the group delay is the group delay relative to the empty air-line,
\begin{equation}
\tau = \frac{L}{c}\left(1 - n - f\frac{dn}{df}\right),
\label{Tg_relative}
\end{equation}
derived from supstracting Eq. (\ref{Tg_cal}) from Eq. (\ref{Tg_empty}).

In the final part of  this section, we describe how to extract the material parameters $\epsilon_r$ and $\mu_r$. A situation of practical interest is the case in which the experimentalist already has some information about the material. For example, if chemical analysis provides additional proof that the material does not contain magnetic elements, one can take $\mu_r =1$ and determine the permittivity from $\epsilon_{r} = n^2$. As we will show later, this leads to better accuracies than the more general method presented below, which requires the determination of $z$.

In many situations however, especially concerning materials under research which contain magnetic elements, magnetic impurities, or magnetic nanoparticles, it is not possible to know beforehand what the electromagnetic properties are. In these situations, one needs to use not only $n$ but also the relative impedance $z$; the material properties are then obtained as
\begin{equation}
\mu_{r} = nz, \epsilon_{r}=n/z, \label{mru}
\end{equation}
where
\begin{equation}
z = \frac{1-\Gamma}{1+\Gamma}. \label{arta}
\end{equation}
The reflection coefficient $\Gamma$ is determined from Eq. (\ref{Gamma2ABC}). But this equation gives $\Gamma$ only up to a sign, since $\pm \Gamma$ both satisfy Eq. (\ref{ABCGamma}). To get the correct sign for $\Gamma$, we have to go back to Eq. (\ref{S11sec2}) or (\ref{S22sec2}) and check which one of $\pm \Gamma$ satisfies them. Note that this does not bring in additional errors, since it is just a sign check.

It is useful to note that even without such a check, only minimal information about the properties of the material may be sufficient. Suppose that $\Gamma$ is the correct solution leading to the correct set of material parameters $\epsilon_{r}$ and $\mu_{r}$. The properties of the conformal mapping, Eq. (\ref{arta}), imply that the opposite-sign solution $-\Gamma$ corresponds to a relative impedance $z^{-1}$. The effect in the final result Eq. (\ref{mru}) is therefore simply to swap the values of permittivity and permeability. In many practical situations, an experienced experimentalist could recognize easily, given two complex numbers and minimal information about the chemical composition of the material, which one is the permittivity and which one is the permeability.

\section{Experiments}
\begin{figure}[!h]
\centering
\includegraphics[width=2.5in]{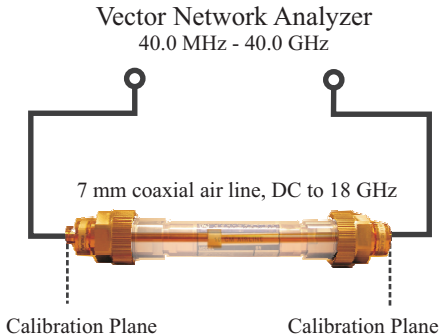}
\caption{The measurement setup used to measure the complex refractive indices of test samples. The 7-mm coaxial air line 
was used as the sample holder to conduct transverse electromagnetic waves. From the diagram, an inserted sample can be 
seen in the middle of the air line.}
\label{ExperimentalSetup}
\end{figure}

To verify the proposed method, the complex refractive indices of polytetrafluoroethylene (PTFE) and polyvinylchloride (PVC) were determined. The reference plane invariant algorithm was tested by measuring the PTFE and PVC sample at various positions relative to the calibration planes of the S-parameter measurements. The group delays and S-parameters were measured by using an Anritsu 37369D vector network analyzer. The measurement setup is shown diagramatically in Fig \ref{ExperimentalSetup}. Transverse electromagnetic waves are transmitted between the measurement ports via a 7mm precision air line set which consists of an outer conductor with beadless connectors and a 3mm center conductor. The air line set was connected to the VNA ports by Anritsu 34ASF50-2 female adapters.

The full 12-term calibration, excluding isolation, was performed prior to the measurements. First the empty air line was measured to obtain $S_{21}^o$ and $\tau_{\rm g}^o$, from which the total length of the air line $L_{\rm air}$ between
the calibration planes was inferred to be 17.3193 cm. Then a toroidal sample was inserted between the inner and the outer
conductor of the air line, and the measurement was repeated again.

\begin{figure}[!h]
\centering
\includegraphics[width=2.5in]{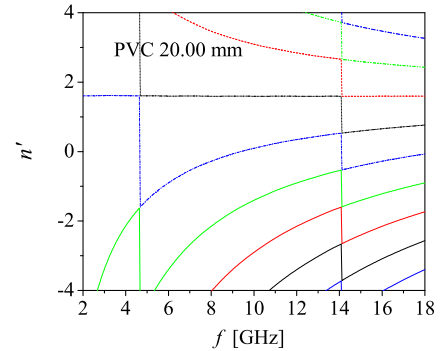}
\caption{Multiple roots of Eq. (\ref{n}) which were used to determine the real part of the refractive index shown in Fig.
\ref{PVC20mm}.}
\label{15RootsPVC20mm}
\end{figure}

\begin{figure}[!h]
\centering
\includegraphics[width=2.5in]{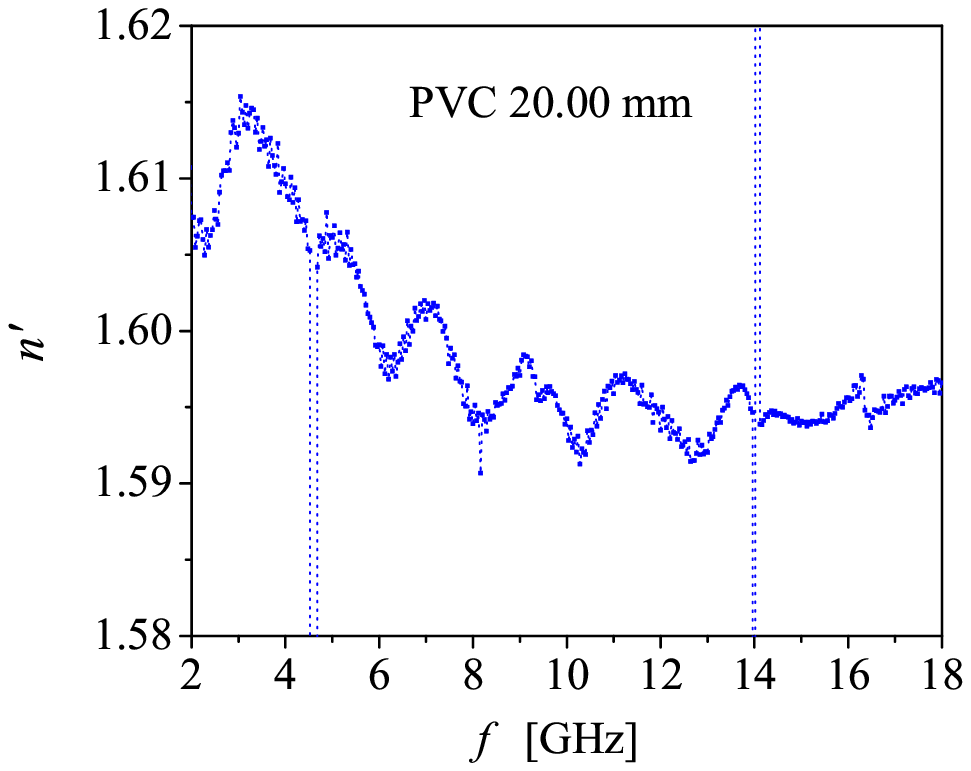}
\includegraphics[width=2.5in]{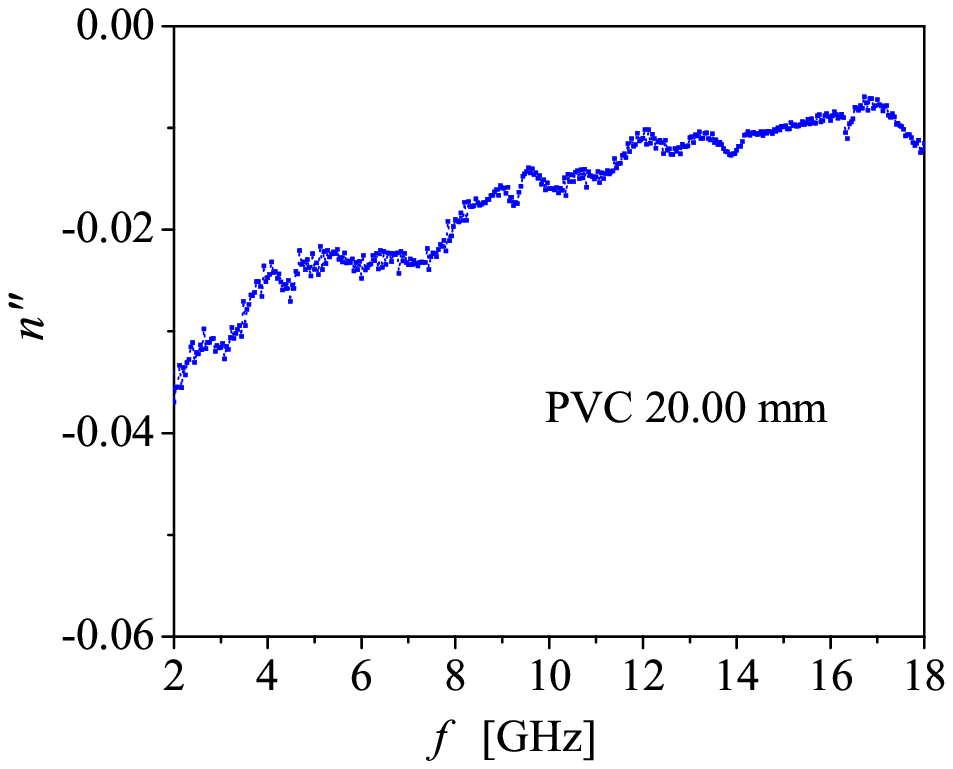}
\caption{The complex refractive index of a 20.00 mm PVC sample determined by using the reference-plane invariant 
method.}
\label{PVC20mm}
\end{figure}

Measurements on a 20.00mm PVC sample give the results shown in Fig. \ref{15RootsPVC20mm} and \ref{PVC20mm}. Within the whole range of measured frequencies (40.0 MHz to 18.0 GHz) there are 450 sampling points. Most of the samplings gave results which are close to the average values. For example, at a frequency of about 12 GHz, the measurement gave $n$ = 1.595 - 0.012 $i$. Comparing the real values, $n^{\prime}$, and the imaginary values, $n^{\prime\prime}$, of the measured refractive index, the algorithm leads to very stable results for the imaginary part, but the real part contains two discontinuities at 4.60 GHz and 14.04 GHz. These frequencies correspond to integer multiples of one-half wavelength inside the sample. By ignoring the discontinuity, the real part of the refractive index can be interpolated as 1.605 around 4.60 GHz. For a wavelength of twice the sample length, this corresponds to a frequency of 4.67 GHz, close to the measured value of 4.60 GHz.

\begin{figure}[!h]
\centering
\includegraphics[width=2.5in]{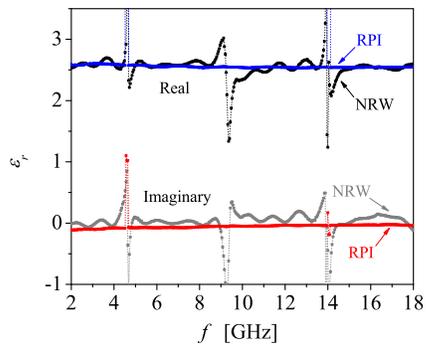}
\caption{Comparison of the complex permittivity obtained using the Nicholson-Ross-Weir method (NRW) and our reference-plane invariant method (RPI) for a PVC sample of length 20.00 mm. The reference-plane positions used in the NRW algorithm are measured by the vector network analyzer.}
\label{NRWRPIPVC20}
\end{figure}

Since PVC is non-magnetic, the relative magnetic permeability is approximately equal to $1$. We can therefore use the measured refractive index to determine the electric permittivity, $\epsilon_r = n^2$. Fig. \ref{NRWRPIPVC20} shows the measured result for the PVC sample of length 20.00 mm obtained by using the proposed method and the Nicholson-Ross-Weir method. We can see that for a non-magnetic material, such as PVC, the complex permittivity can be measured using this method with relatively high accuracy.
\begin{figure}[!h]
\centering
\includegraphics[width=2.5in]{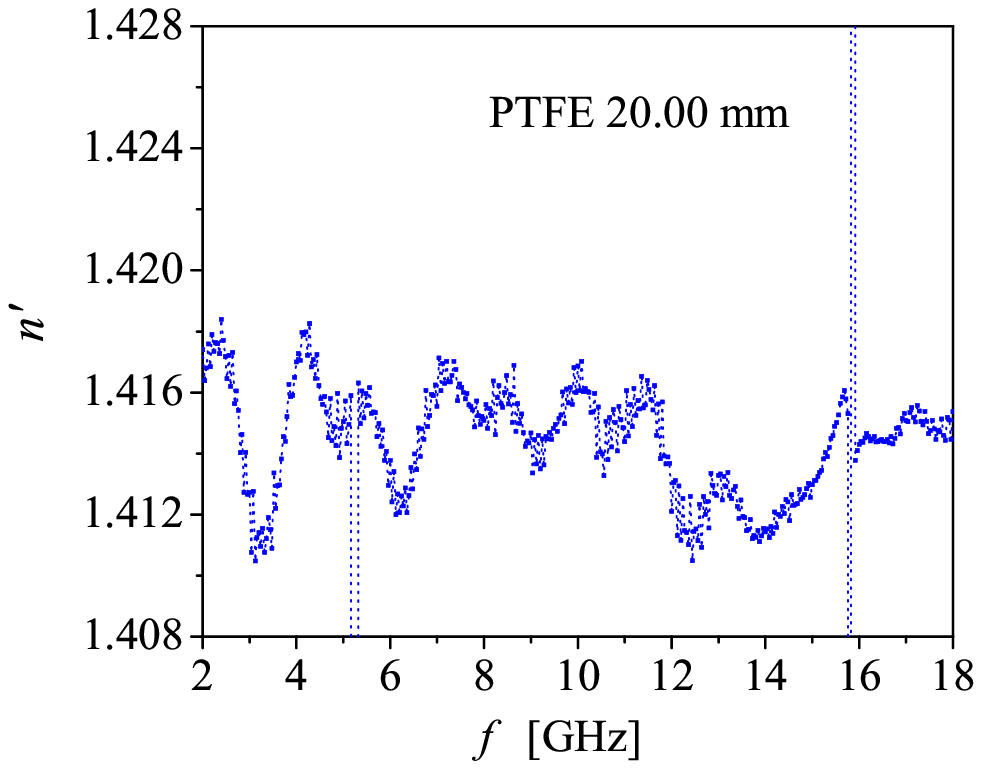}
\includegraphics[width=2.5in]{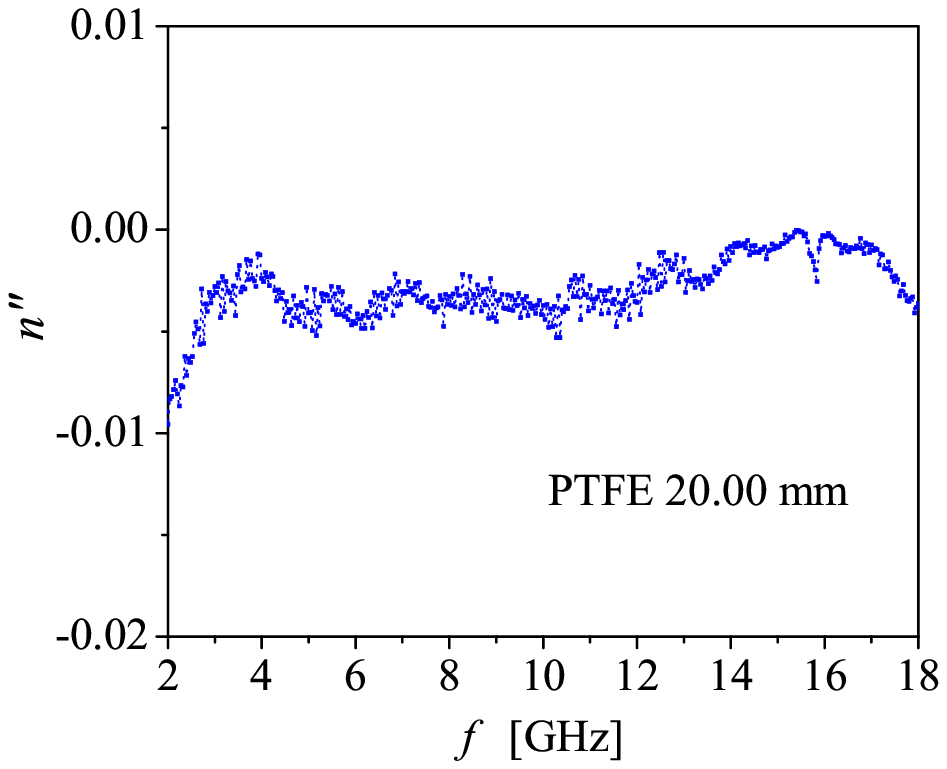}
\caption{The complex refractive index of a 20.00 mm PTFE sample determined by using the reference-plane invariant method.}
\label{PTFE20mm}
\end{figure}
\begin{figure}[!h]
\centering
\includegraphics[width=2.5in]{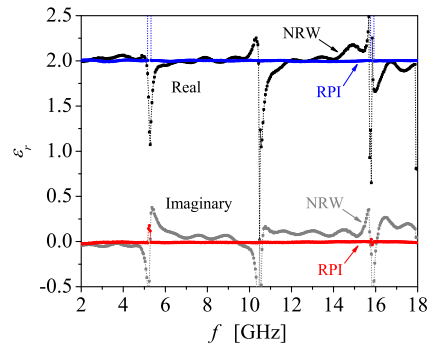}
\caption{Comparison of the complex permittivity obtained using the Nicholson-Ross-Weir method (NRW) and our reference-plane invariant method (RPI) for a PTFE sample of length 20.00 mm. The reference-plane positions used in the NRW algorithm are measured by the vector network analyzer.}
\label{NRWRPIPTFE20}
\end{figure}

To further examine and verify the proposed algorithm, we also performed a  measurement on a PTFE sample of length 20.00 mm. The results are shown in Fig. \ref{PTFE20mm}. At 10 GHz, the index of refraction has the value of about 1.415 - 0.004$i$.
The spectrum of the real values, $n^\prime$, is discontinuous at 5.24 and 15.84 GHz, while the spectrum of the imaginary values, $n^{\prime\prime}$, is relatively stable over the whole range of frequency, but we can see one relatively large 
peak near 15.84 GHz. In Fig. \ref{NRWRPIPTFE20}, the electric permittivity of PTFE is determined from $\epsilon_r = n^2$. The spectra obtained by the new method show much less errors compared to the ones from the NRW algorithm, especially at higher frequencies where we have larger uncertainties in  $L_1$ and $L_2$.

\begin{figure}[!h]
\centering
\includegraphics[width=2.5in]{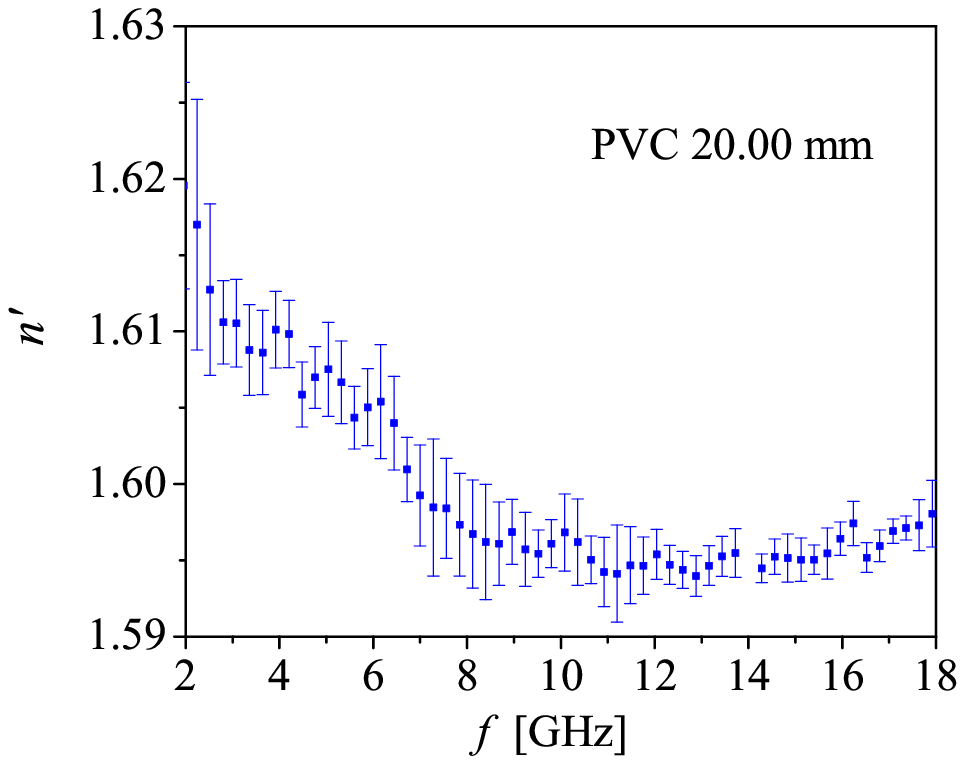}
\includegraphics[width=2.5in]{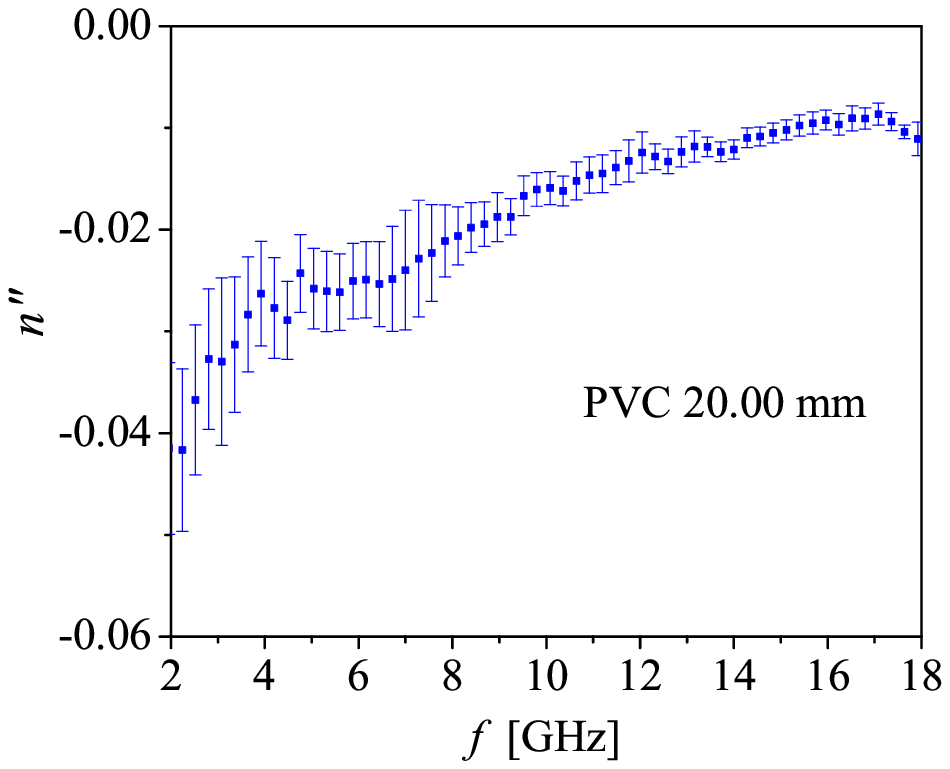}
\caption{The refractive index of PVC measured by using a sample of length 20.00 mm.}
\label{MeanSDPVC20mm}
\end{figure}
\begin{figure}[!h]
\centering
\includegraphics[width=2.5in]{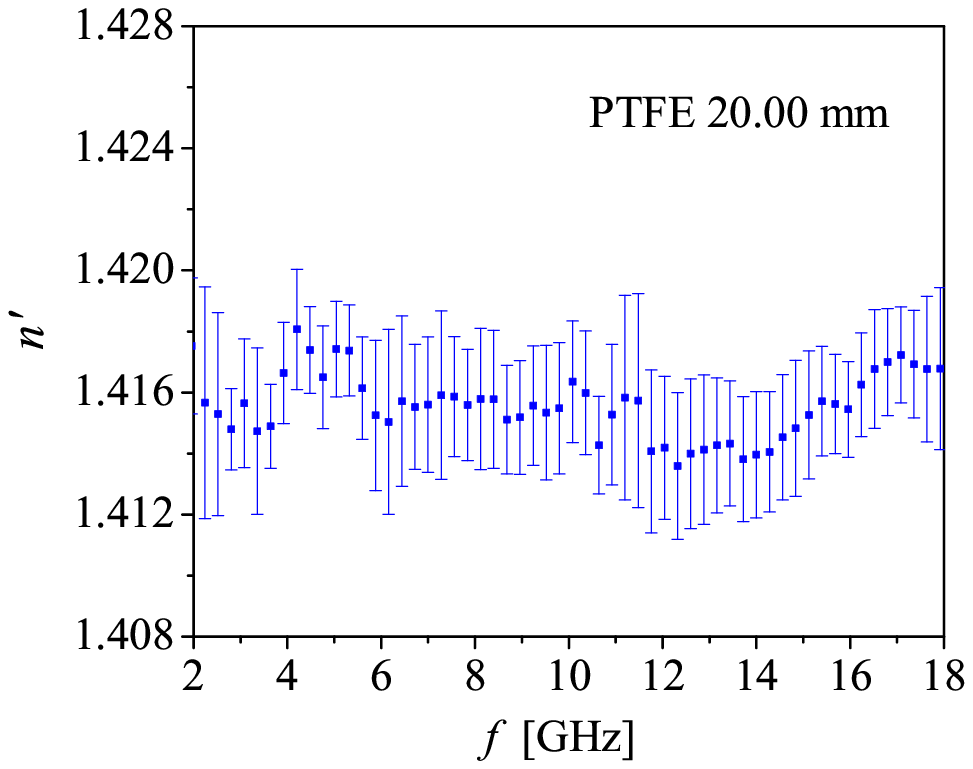}
\includegraphics[width=2.5in]{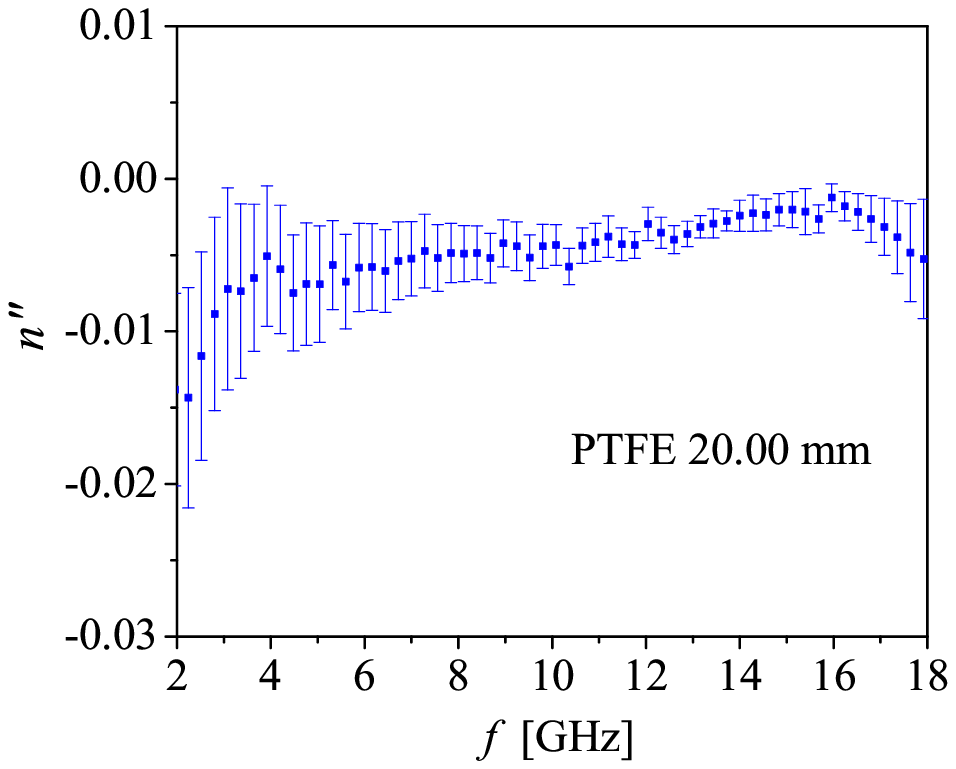}
\caption{The refractive index of PTFE measured by using a sample of length 20.00 mm.}
\label{MeanSDPTFE20mm}
\end{figure}

The method is invariant with respect to the position of the reference-planes, therefore, theoretically, the position of the sample inside the air line should not affect the results. To verify this statement, a number of measurements were repeated by placing the samples at different positions inside the air line. The statistical analysis of the measured results are shown graphically in Fig. \ref{MeanSDPVC20mm}-\ref{MeanSDPTFE20mm}. The error bars represent the uncertainties, i.e. standard deviations, of the measurements. It can be seen that the uncertainties of either $n^\prime$ or $n^{\prime\prime}$ are lower at higher frequencies.

For PTFE, the uncertainties of the real values vary slowly with frequency, while the uncertainties of the imaginary values are lower at higher frequencies and become larger again near 18 GHz. This might be explained in terms of energy losses within the line because the measurement on the empty line shows higher losses at frequencies near 18 GHz. In PVC, this effect is not observed because the material losses are larger and tend to dominate the air line imperfections.

\section{Error Analysis}

Using the proposed method, the complex refractive index is determined by measuring the transmission/reflection signals, the group delay, the air line length and the sample length. According to the rules of error propagation, higher errors in the sample length and S-parameter measurements will lead to higher uncertainties in the measured refractive index. Experimentally, the transmission and reflection signals are measured by a VNA, therefore the accuracy of the measurement is limited somehow by the VNA uncertainties. In fact, there are also other causes of errors to consider, such as the imperfections of the air line and the air gaps between the surfaces of the sample and the line. The errors due to the air gaps can be compensated by mathematical models. Equations for the air gap correction can be found in the literature \cite{BakerJarvisNIST}, \cite{BusseyGray1962}.

The analysis is simplified by assuming that the errors due to air gaps, connector mismatches and air line losses are small compared to the errors due to the sample length and S-parameter measurements. The uncertainties of reference-plane positions are also neglected because the final equations are theoretically reference-plane invariant.

Experimentally, the S-parameters are independently measured by a VNA, so if the errors of the magnitudes and phases of the S-parameters are assumed to be independent, the relative uncertainty can be calculated using the rules of error propagation as follows:
\begin{equation}
\hspace{-0.2in}\frac{\delta n}{n}=\frac{1}{n}\sqrt{\left(\frac{\partial n}{\partial L} \delta L 
\right)^2+\sum_{\alpha}\left[\left(\frac{\partial n}{\partial \left| S_{\alpha} \right|} \delta \left| S_{\alpha}\right| 
\right)^2 + \left( \frac{\partial n}{\partial \theta_{\alpha}}\delta\theta_{\alpha}\right)^2\right]}.
\label{RelativeerrorRIdef}
\end{equation}
The index $\alpha\in \{11,12,21,22,21o\}$ runs over the scattering parameters, $S_{\alpha}=\left| 
S_{\alpha}\right|\exp(i\theta_{\alpha})$. (The index $21o$ refers to the transmission through the empty air line.) Eq. 
(\ref{RelativeerrorRIdef}) can be further expressed as
\begin{eqnarray}
\hspace{-0.2in}
\left(\frac{\delta n}{n}\right)^2=\left(-\frac{1}{L} + \frac{1}{P\ln P}\frac{\partial P}{\partial L}\right) (\delta 
L)^2\hspace{1cm}{\rm 
\hspace{0.5cm}}&&\label{RelativeerrorRI}\\&&\nonumber\\+\frac{1}{P^{2}\ln^{2}P}\sum_{\alpha}\left[\left(\frac{\partial 
P}{\partial \theta_{\alpha}}\right)^{2}(\delta\theta_{\alpha})^2 +\left(\frac{\partial P}{\partial 
\left|S_{\alpha}\right|}\right)^{2}(\delta \left|S_{\alpha}\right|)^2\right].&&\nonumber
\end{eqnarray}

\begin{figure}[!h]
\centering
\includegraphics[width=2.5in]{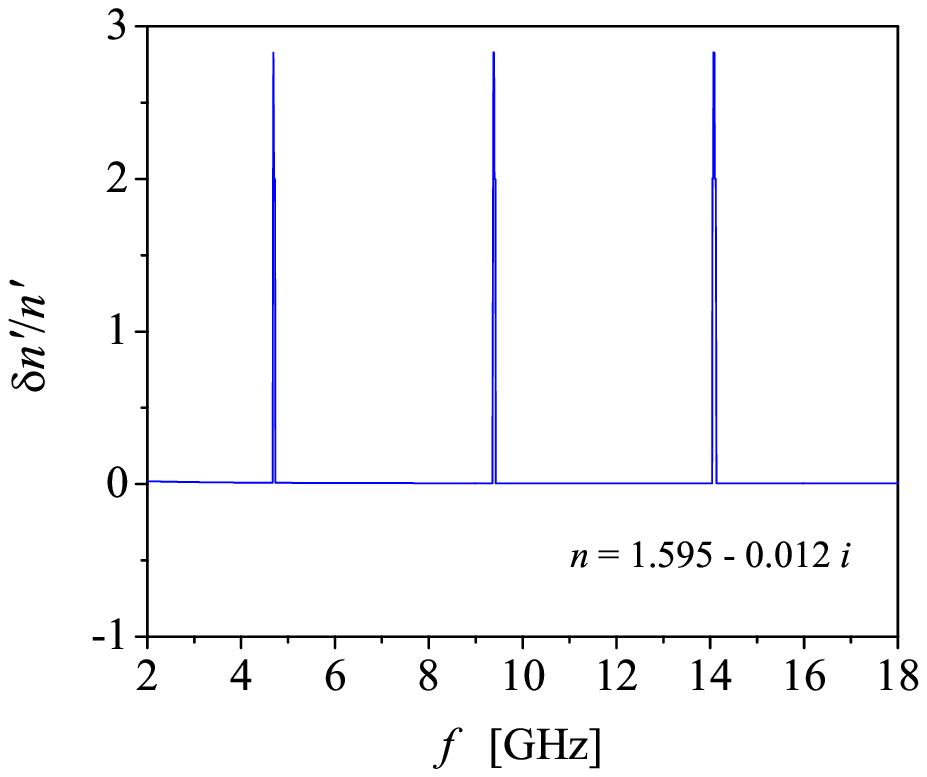}
\includegraphics[width=2.5in]{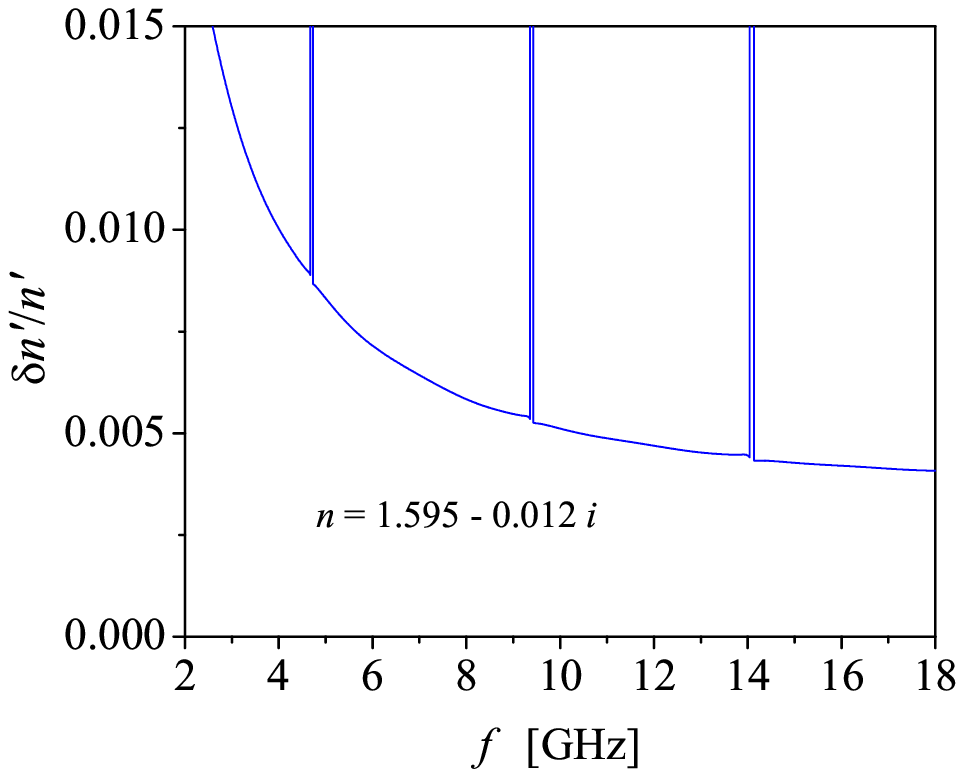}
\includegraphics[width=2.5in]{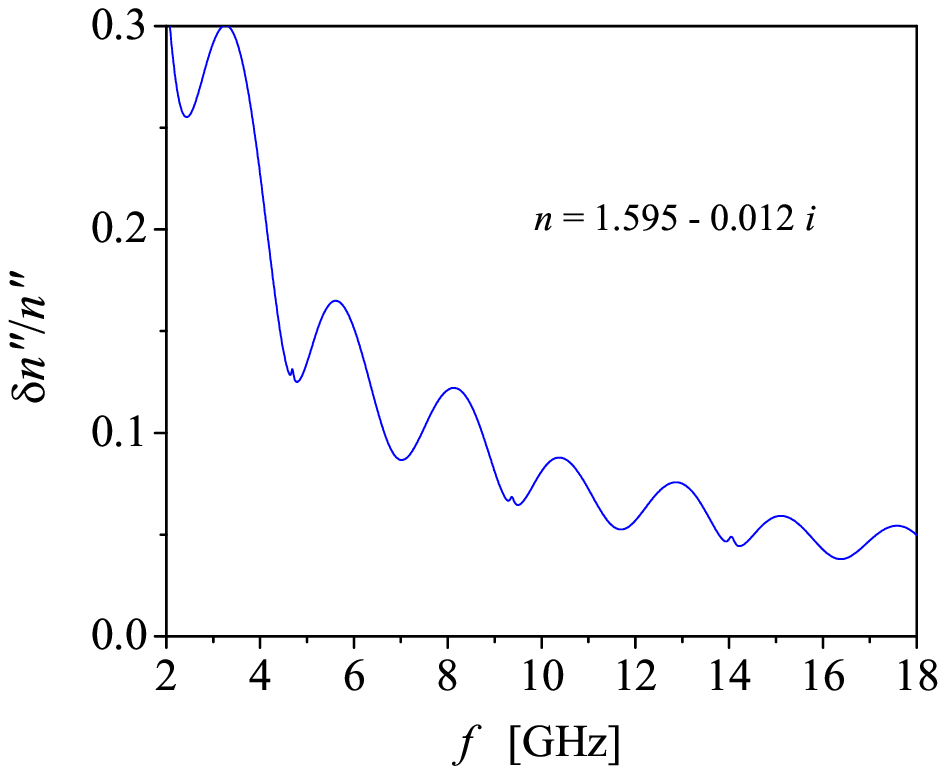}
\caption{The relative uncertainty of refractive index for a model sample of length 20.00 mm and $n$ = 1.595 - 0.012$i$. 
The middle graph shows the same result as the upper graph at a magnified scale. The relative uncertainties of the real 
part, $\delta n^\prime / n^\prime$, are very high around the resonant frequencies, but close to zero elsewhere. Both 
the real and imaginary spectra show relatively low uncertainties at high frequencies.}
\label{PVCuncertaintySimulation}
\end{figure}

\begin{figure}[!h]
\centering
\includegraphics[width=2.5in]{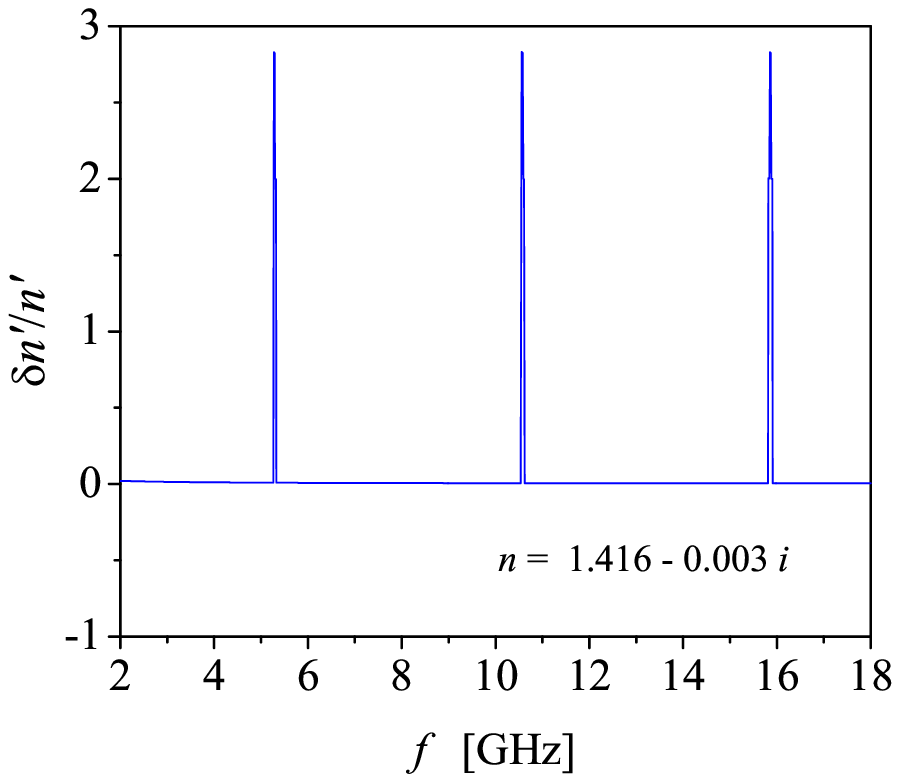}
\includegraphics[width=2.5in]{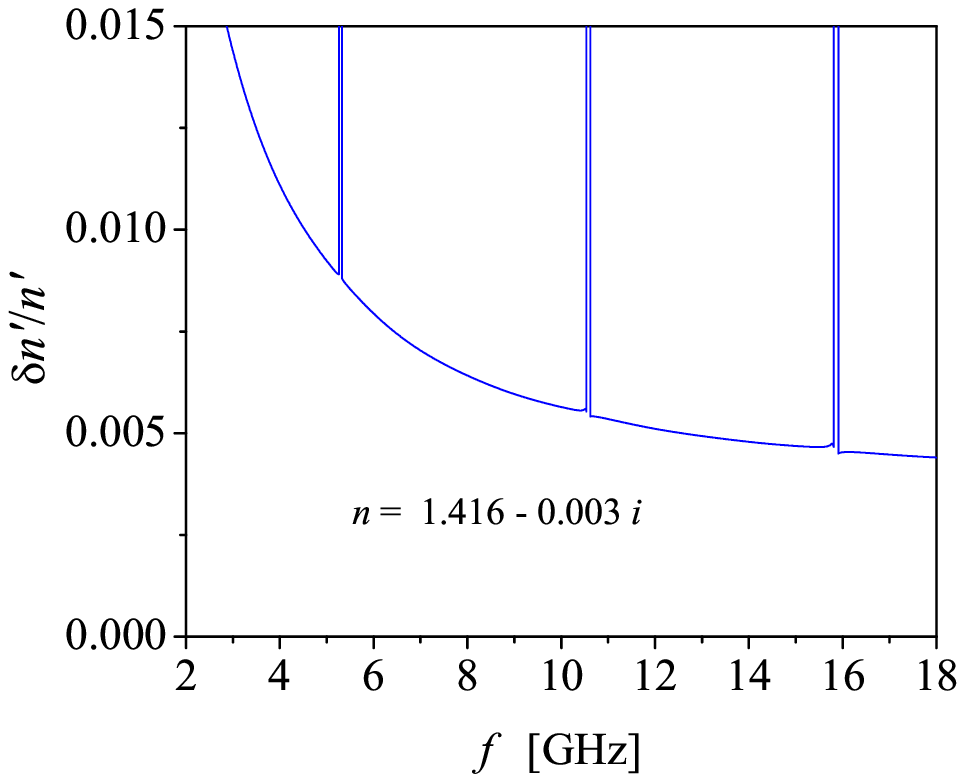}
\includegraphics[width=2.5in]{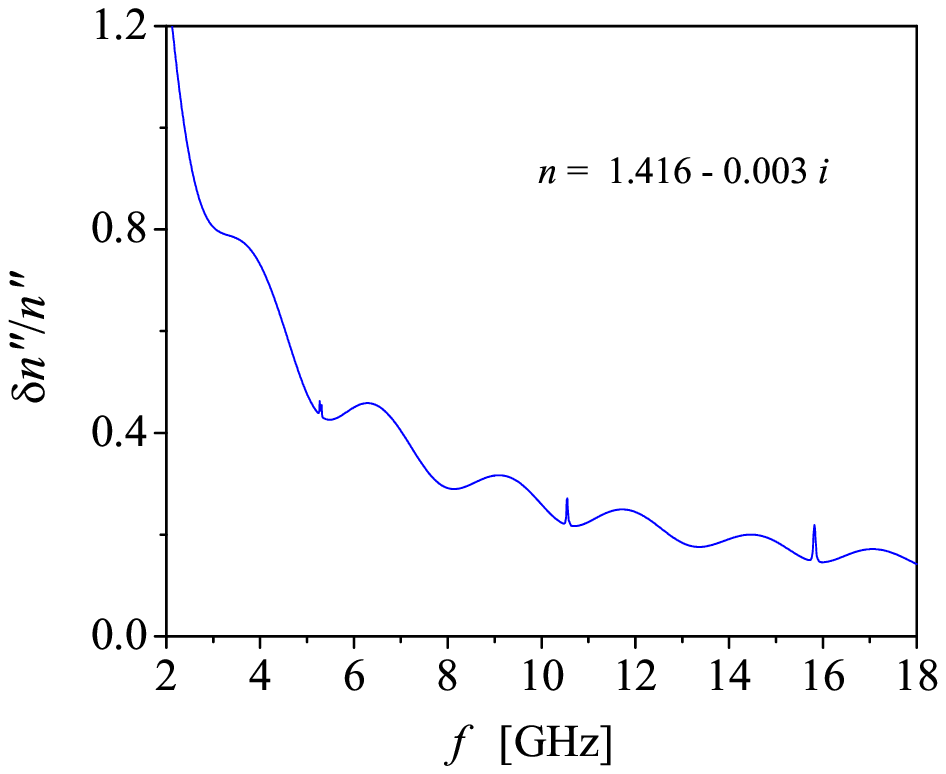}
\caption{The relative uncertainty of refractive index for a model sample of length 20.00 mm and $n$ = 1.416 - 0.003$i$. 
The middle graph shows the same result as the upper graph at a magnified scale. The imaginary spectrum shows some small 
peaks at resonant frequencies.}
\label{PTFEuncertaintySimulation}
\end{figure}

In the remaining part of this section, the errors involved in our experimental determination of the dielectric properties of PVC and PTFE will be analyzed. The derivatives in (\ref{RelativeerrorRI}) can be analytically calculated, but the results are complicated and it is in fact simpler to numerically calculate the relative uncertainties in (\ref{RelativeerrorRIdef}). Fig. \ref{PVCuncertaintySimulation} and Fig. \ref{PTFEuncertaintySimulation} shows the numerical results for two models, one with $n$ = 1.595 - 0.012$i$ and another with $n$ = 1.416 - 0.003$i$.

\begin{figure}[!h]
\centering
\includegraphics[width=2.5in]{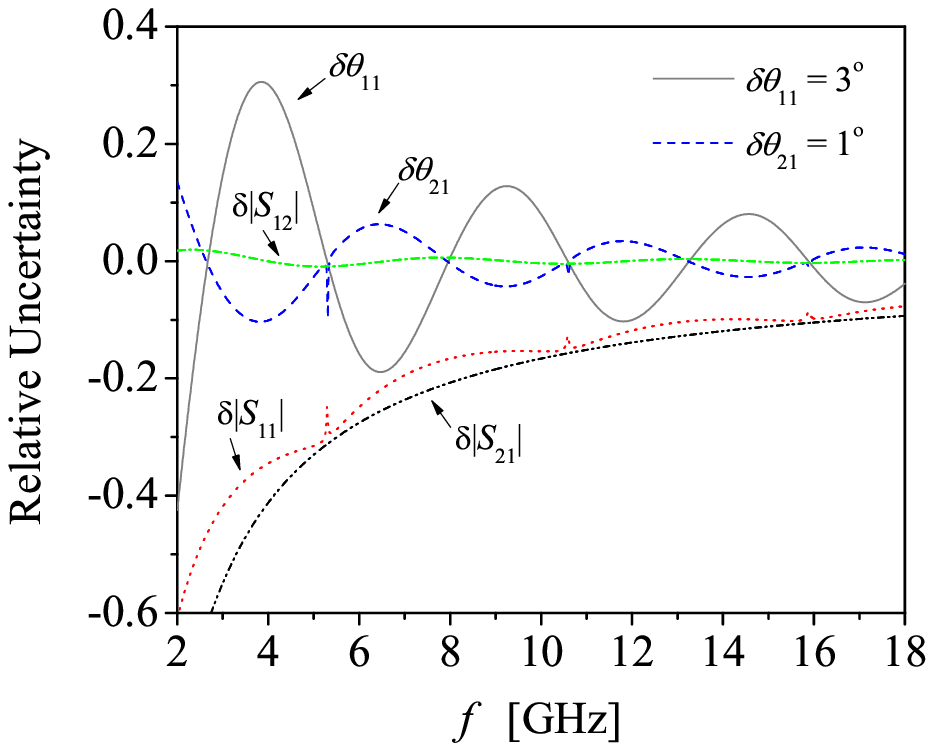}
\includegraphics[width=2.5in]{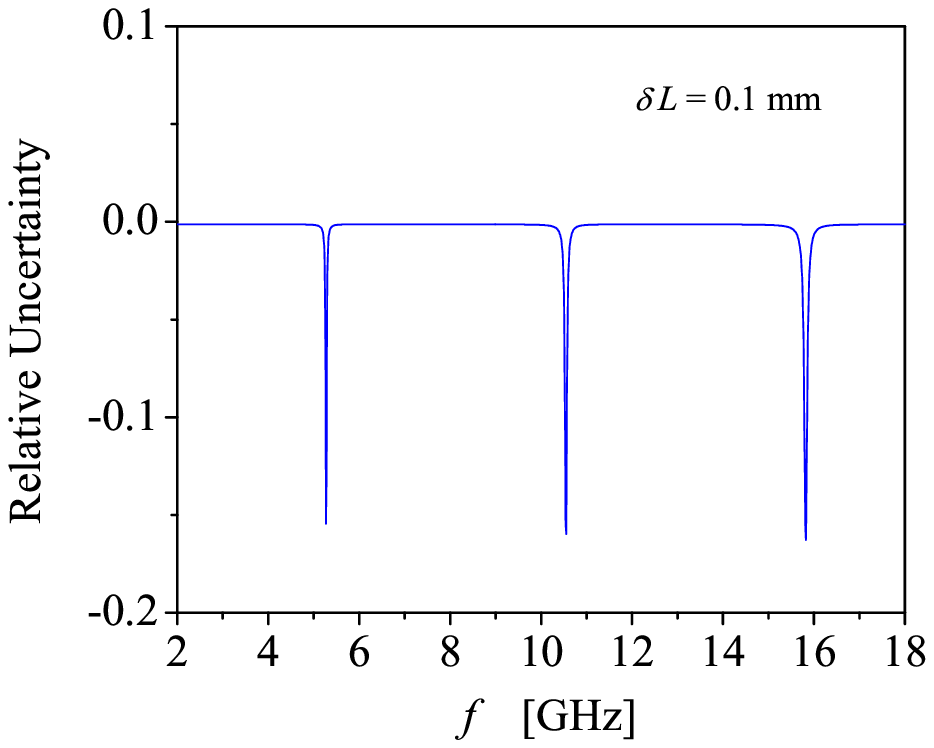}
\caption{The relative uncertainty of $n^{\prime\prime}$ for a model sample of length 20.00 mm and $n$ = 1.416 - 
0.003$i$. Each line represents relative uncertainty, $(1/n^{\prime\prime})\times (\delta n^{\prime\prime}/\delta x)\times \delta x$, caused by a single-source error, $\delta x$. The upper graph shows the relative uncertainty due to the S-parameter errors, i.e. $x \in \lbrace\theta_{11}, \theta_{21}, |S_{11}|, |S_{12}|, |S_{21}|\rbrace$. The lower graph shows the uncertainty due to the sample length error.}
\label{PTFEuncertaintyImagSim}
\end{figure}

The numerical calculations are done based on the assumption that the materials are non-magnetic. That is, the
relative permeability of each model is taken to be 1, $\mu_r = 1$, and the relative permittivity is equal to the square of the refractive index, $\epsilon_r = n^2$. The errors due to the sample length and S-parameter measurements are set as
follows: $\delta L = 0.1$ mm, $\delta |S_{\alpha}| = 0.002$, $\delta \theta_{11} = \delta \theta_{22} = 3^\circ$ and 
$\delta \theta_{21} =  \delta \theta_{12} = \delta \theta_{21o} = 1^\circ$. The error $\delta L$ is set greater than the actual value in order to investigate the behavior of the corresponding uncertainty. In the experiment, the PVC and PTFE samples were precisely machined, and the uncertainties of the lengths and diameters of the samples are less than 0.04 mm.

It can be seen from the relative uncertainties in Fig. \ref{PVCuncertaintySimulation} and the experimental results in Fig. \ref{PVC20mm} and Fig. \ref{MeanSDPVC20mm} that the errors determined theoretically correspond well with the errors determined experimentally. As in other methods based on transmission/reflection measurements, uncertainties are lower at higher frequency. This is understandable, because a significant change in the phase due to the presence of the material in the air line requires that the sample length should not be too small compared to the wavelength. Around  4.70, 9.39 and 14.08 GHz, the simulation shows discontinuities in the real part spectrum. As mentioned before, the origin of these discontinuities is of physical nature, namely the occurrence of Fabry-P\'erot resonances in the sample.
However, our results for $n$ (and also for $\epsilon_r = n^2$ in the case of non-magnetic materials) are less sensitive to this effect than those obtained {\it via} the original NRW algorithm: this can be seen already in Fig. \ref{NRWRPIPVC20}. The  difference is mathematical: in the NRW algorithm the discontinuity is generated
by the relative impedance $z=(1-\Gamma)/(1+\Gamma)$ \cite{Boughriet}, while for non-magnetic materials our algorithm  avoids calculating this quantity. A comparison with  the situation in which $z$ needs to be calculated is given also below in Section \ref{cons}.

Fig. \ref{PTFEuncertaintySimulation} shows the simulated results for a sample  with $n = 1.416-0.003i$. It is clear from the simulation that the relative uncertainties in $n^\prime$ and $n^{\prime\prime}$ are again prone to be higher at lower frequencies. The uncertainty in $n^{\prime\prime}$ oscillates in frequency domain and has small peaks at resonant frequencies, {\it i.e.} frequencies corresponding to integer multiples of one-half wavelength inside the sample.

Fig. \ref{PTFEuncertaintyImagSim} shows a comparison between the uncertainties due to each source of error. The uncertainties originating from the S-parameter errors oscillate and decrease in amplitude when the frequency is increased. This explains why the total relative uncertainties of $n^{\prime\prime}$ shown in Fig. \ref{PTFEuncertaintySimulation} behave in a similar way. The error in the sample length measurement leads to high uncertainties at frequencies corresponding to integer multiples of one-half wavelength, but very low uncertainties at other frequencies. In experiments, sample lengths can be measured with high accuracy, so the corresponding errors are much smaller than the simulated results.

We have also simulated the relative uncertainties of  $n^\prime$ and $n^{\prime\prime}$ caused by
errors in the air-line length, and found that the corresponding uncertainties behave similarly to those caused by the sample-length error. Experimentally, the error due to the air-line length depends on the accuracy of the TRU calibration.

\section{Theoretical considerations on measurement errors for various types of materials}
\label{cons}

For dielectric materials, we have already analyzed measured data for PVC and PTFE; for magnetic materials, no reliable standards or reference materials have emerged. In this section, we will give a systematic analysis of measurement errors for both dielectric and magnetic materials based on simulated experiments. First, a set of S-parameters and group delays are generated based on given values of the sample length $L$, the air-line length $L_{air}$, the complex electric permittivity $\epsilon_r$ and the complex magnetic permeability $\mu_r$. Then errors are added to these data and the results are used as the inputs of the reference-plane-invariant algorithm.

\subsection{Low-loss dielectric materials}
\begin{figure}[!h]
\centering
\includegraphics[width=2.7in]{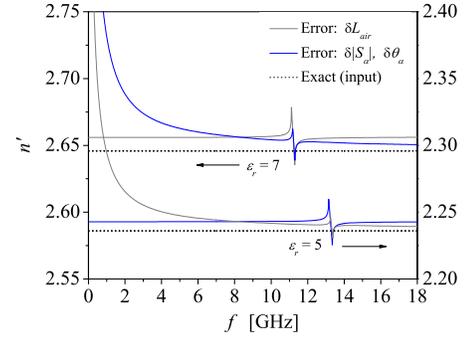}
\caption{Simulated real-part spectra of the refractive indices for dielectric materials with $\epsilon_r$ $=$ 5 and 7.}
\label{DielectricLowLoss_Ren}
\end{figure}

\begin{figure}[!h]
\centering
\includegraphics[width=2.5in]{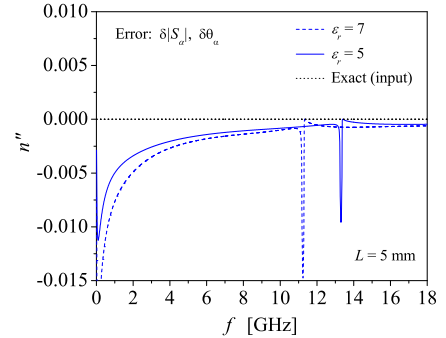}
\caption{Simulated imaginary-part spectra of the refractive indices for dielectric materials with $\epsilon_r$ $=$ 5 and 7.}
\label{DielectricLowLoss_Imn}
\end{figure}

Figs. \ref{DielectricLowLoss_Ren}-\ref{DielectricLowLoss_Imn} show the results of simulated measurements of the refractive index for two zero-loss dielectric materials ($\epsilon_r$ $=$ 5 and 7) with sources of errors $\delta|S_\alpha|$ $=$ $-0.001$, $\delta\theta_\alpha$ $=$ $1^\circ$ and $\delta L_{air}$ $=$ $0.1$ mm. The length of the sample was taken $L=5$ mm.

Similarly to what we have shown in Fig. \ref{PTFEuncertaintyImagSim}, we see clearly that when the sample length is small compared to the wavelength, the errors become significantly higher at lower frequencies. We also check the code by setting $\delta|S_\alpha|$ $=$ $0$, $\delta\theta_\alpha$ $=$ $0^\circ$, $\delta L_{air}$ $=$ $0$, and find that we recover the exact initial values of $\epsilon_{r}$.

\begin{figure}[!h]
\centering
\includegraphics[width=2.5in]{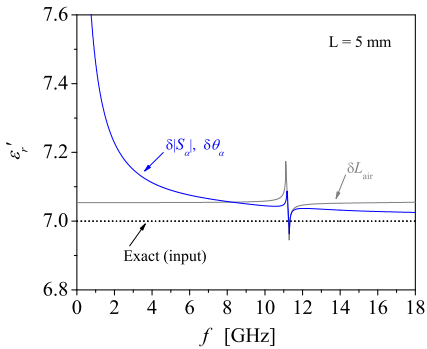}
\caption{Simulated real spectra of electric permittivity for a dielectric material with $\epsilon_r$ $=$ $7$, calculated from the relation: $\epsilon_r = n^2$. The errors are set as follows: $\delta |S_{\alpha}| = -0.001$, $\delta \theta_{\alpha} = -1^o$ and $\delta L_{air} = 0.1$ mm.}
\label{DielectricLowLoss_e-from-n}
\end{figure}

\begin{figure}[!h]
\centering
\includegraphics[width=2.5in]{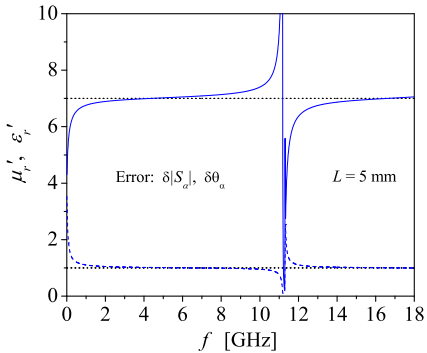}
\includegraphics[width=2.5in]{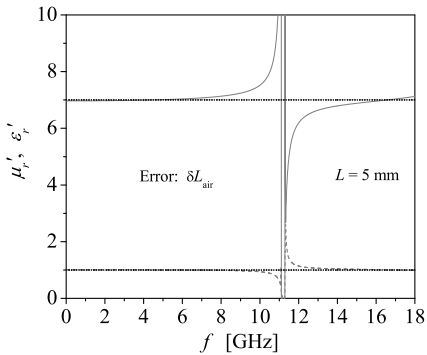}
\caption{Simulated real spectra of electric permittivity and magnetic permeability (extracted as $\epsilon_r = n/z$ and $\mu_r = nz$) for a dielectric material with $\epsilon_r$ $=$ $7$, with the errors in the S-parameters (upper plot) and $L_{air}$ (lower plot).}
\label{DielectricLowLoss-e-u}
\end{figure}

In Fig. \ref{DielectricLowLoss_e-from-n} we present the permittivity spectra determined from the complex refractive index using $\epsilon_r = n^2$, and in Fig. \ref{DielectricLowLoss-e-u} we show the same spectra obtained using the version of the algorithm in which $\epsilon_r$ and $\mu_r$ are extracted {\it via} the relative impedance $z$, $\epsilon_r = n/z$ and $\mu_r = nz$. In the second case, the errors around the resonant frequencies get larger. The mathematical origin of these errors is similar to that of the NRW algorithm: measurement errors become magnified at and around the frequencies at which the reflection goes to zero. The difference is that here we do not have errors due to the reference-plane positions.

\subsection{Lossy dielectric materials}
For lossy dielectric materials, resonances do not lead to zero reflected signals at the resonant frequencies, which results in smaller errors in the permittivity and permeability spectra.

\begin{figure}[!h]
\centering
\includegraphics[width=2.5in]{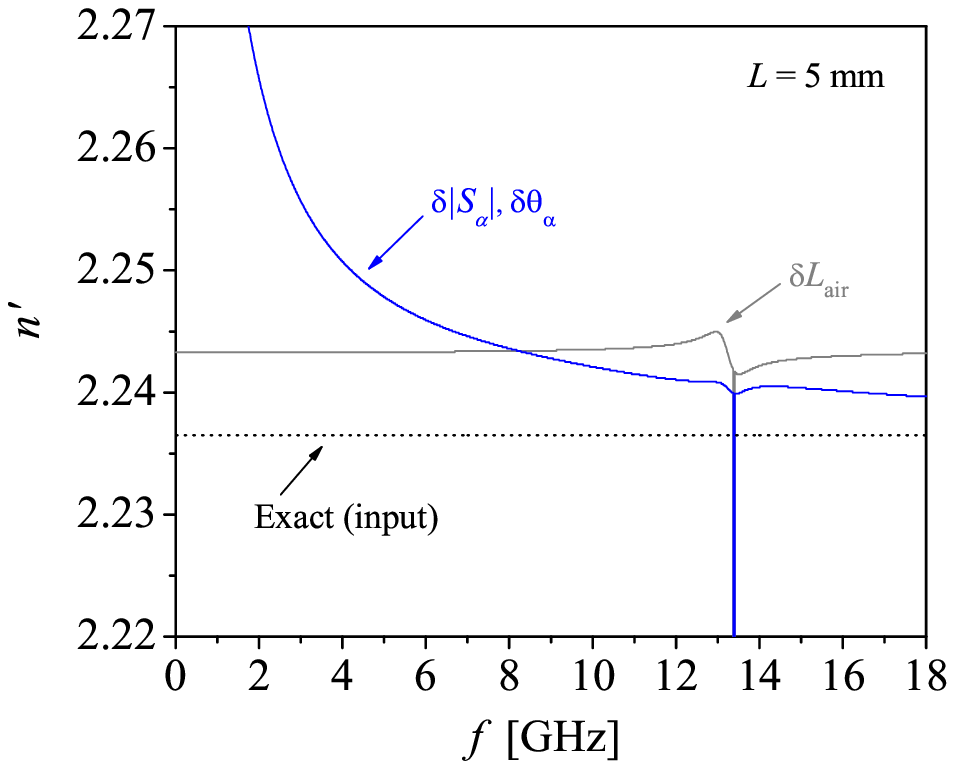}
\includegraphics[width=2.5in]{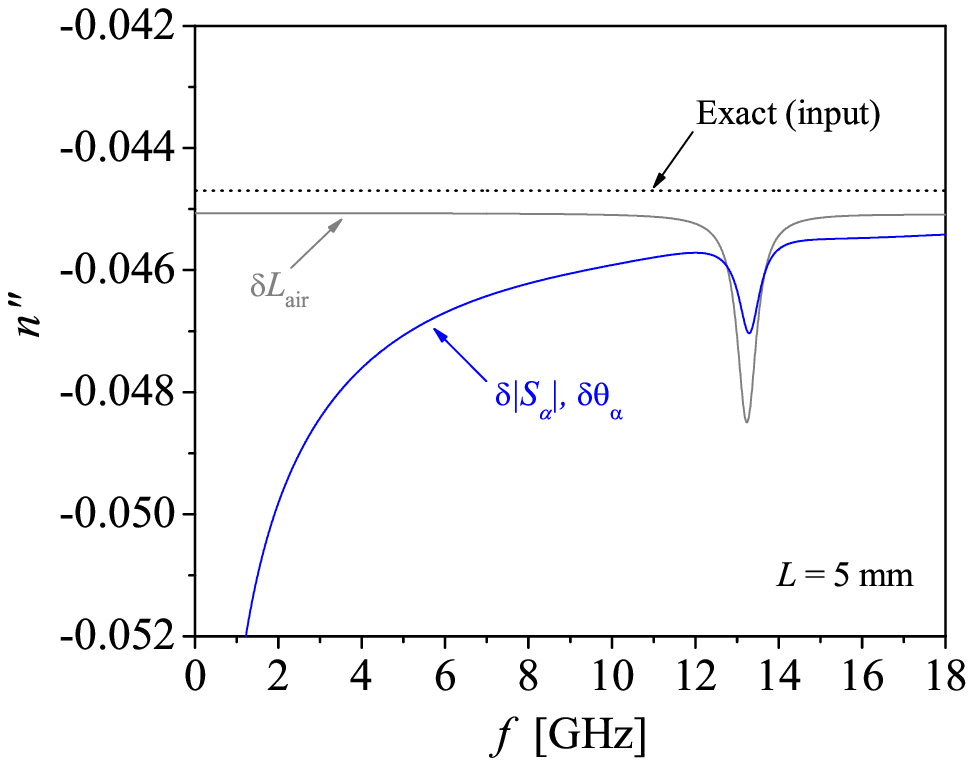}
\caption{The simulated spectra of the complex refractive indices for a lossy dielectric material with $\epsilon_r$ $=$ $5-0.2i$, $\mu_r$ $=$ $1$. The errors are set as follows: $\delta |S_{\alpha}| = -0.001$, $\delta \theta_{\alpha} = -1^o$ and $\delta L_{air} = 0.1$ mm.}
\label{DielectricHighLoss_n}
\end{figure}

\begin{figure}[!h]
\centering
\includegraphics[width=2.5in]{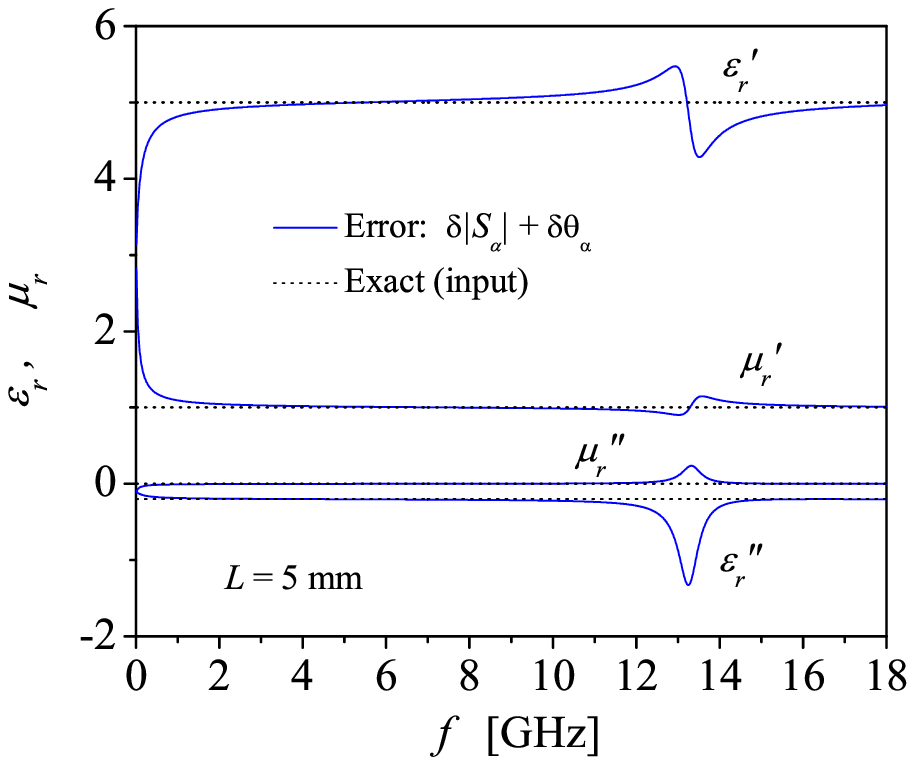}
\includegraphics[width=2.5in]{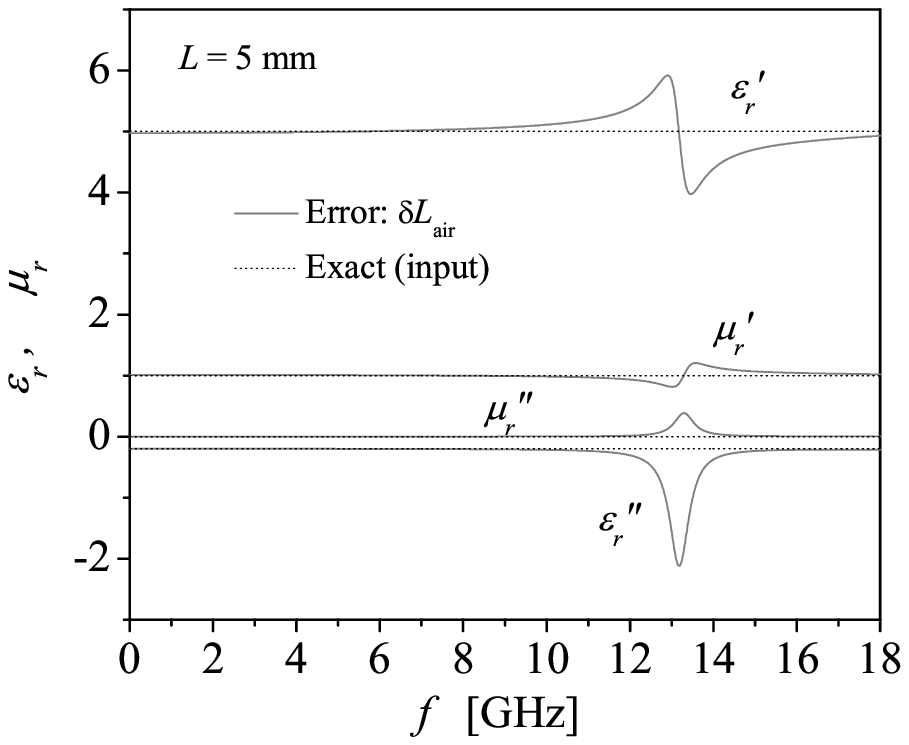}
\caption{The simulated spectra of the complex permittivity and permeability for a lossy dielectric material with $\epsilon_r$ $=$ $5-0.2i$, $\mu_r$ $=$ $1$. The upper graph presents the errors due to S-parameters, with $\delta |S_\alpha| = -0.001$ and $\delta \theta_\alpha = -1^o$, the lower graph is simulated with $\delta L_{air} = 0.1$ mm.}
\label{DielectricHighLoss_e_u}
\end{figure}

To quantify how the energy loss affects the errors, the S-parameters generated for $\epsilon_r$ $=$ $5 - 0.2i$ are used as the input of the reference-plane invariant algorithm. The simulation is done with $\delta|S_\alpha|$ $=$ $-0.001$, $\delta\theta_\alpha$ $=$ $-1^\circ$ and $\delta L_{air}$ $=$ $0.1$ mm. The results are shown in Fig. \ref{DielectricHighLoss_n}, indicating that the errors are quite low over the entire frequency range. It is now possible to extract, with small errors, $\epsilon_{r}$ by using $\epsilon_r = n^2$.

Alternatively, one can extract the complex $\epsilon_r$ using the relative impedance $z$, with  results shown in Fig. \ref{DielectricHighLoss_e_u}. Similar to zero-loss materials, measurement uncertainties cause significant errors at and around Fabry-P\'erot resonant frequency, {\it i.e.} frequency corresponding to integer-multiple of one-half wavelength inside the sample. However, for lossy materials, only finite peaks are found, with no divergence, in both spectra of $\epsilon_r$ and $\mu_r$.

\subsection{Magnetic materials}
Next we consider the case of magnetic materials, {\it i.e.} $\mu_r$ $\neq$ $1$. In general, these materials are lossy and $\mu'$
has relatively low values.

\begin{figure}[!h]
\centering
\includegraphics[width=2.5in]{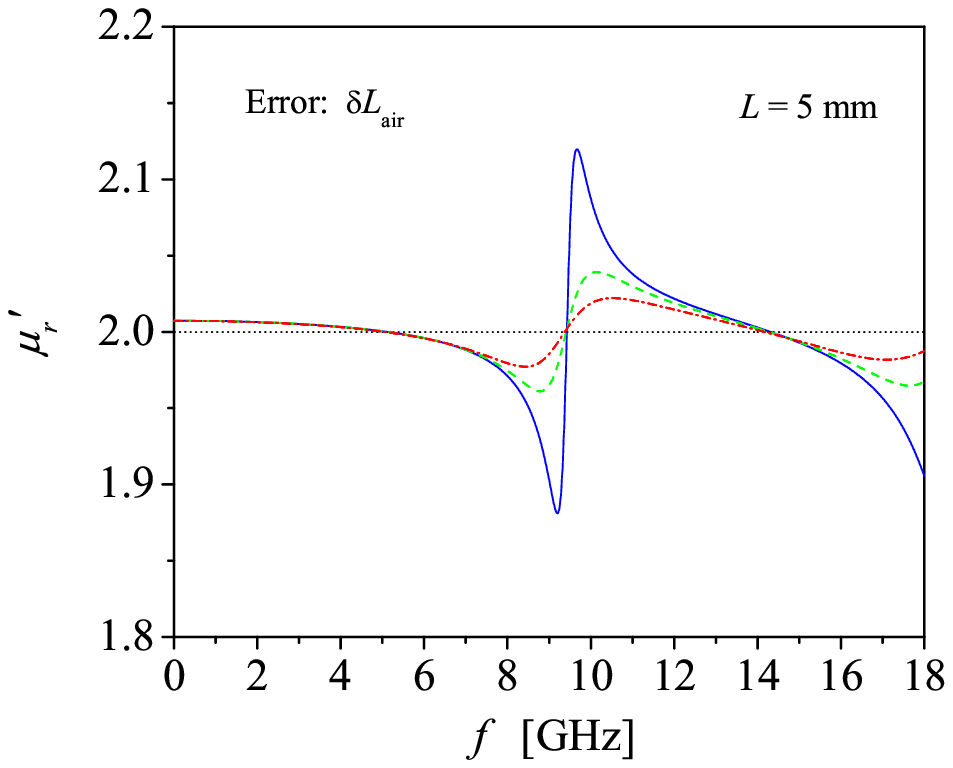}
\includegraphics[width=2.5in]{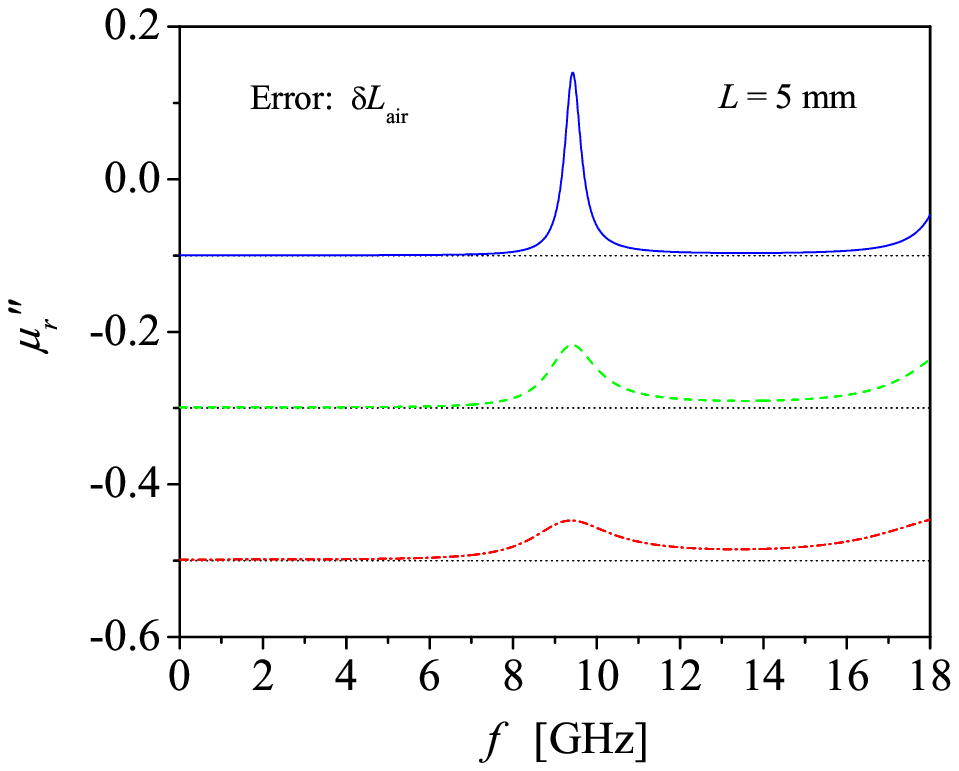}
\includegraphics[width=2.5in]{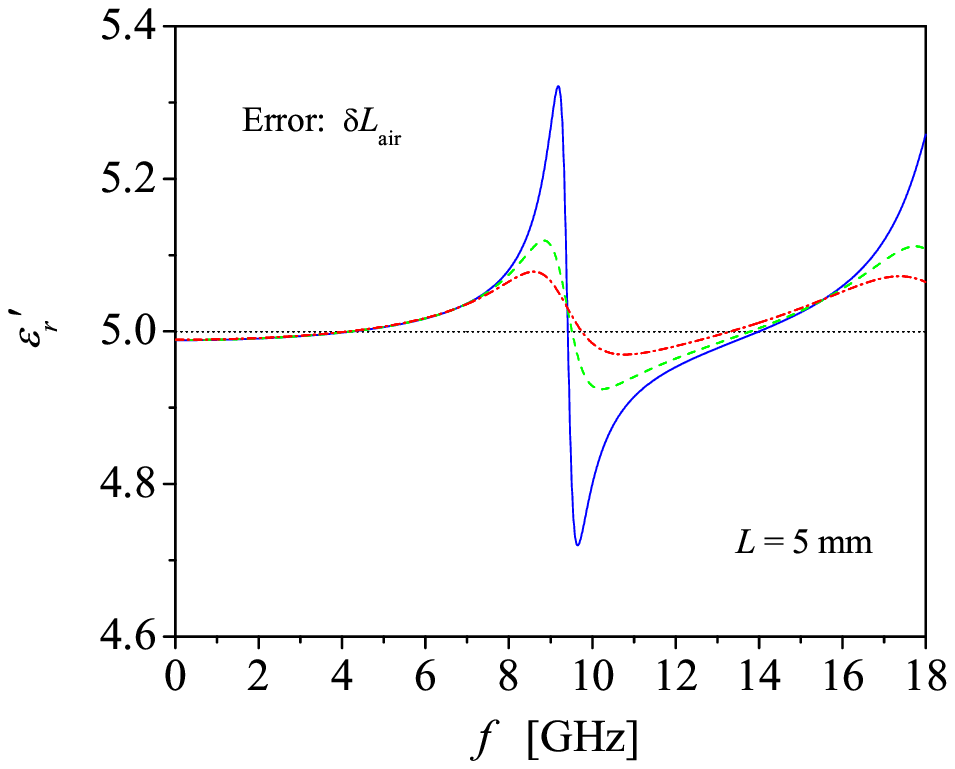}
\includegraphics[width=2.5in]{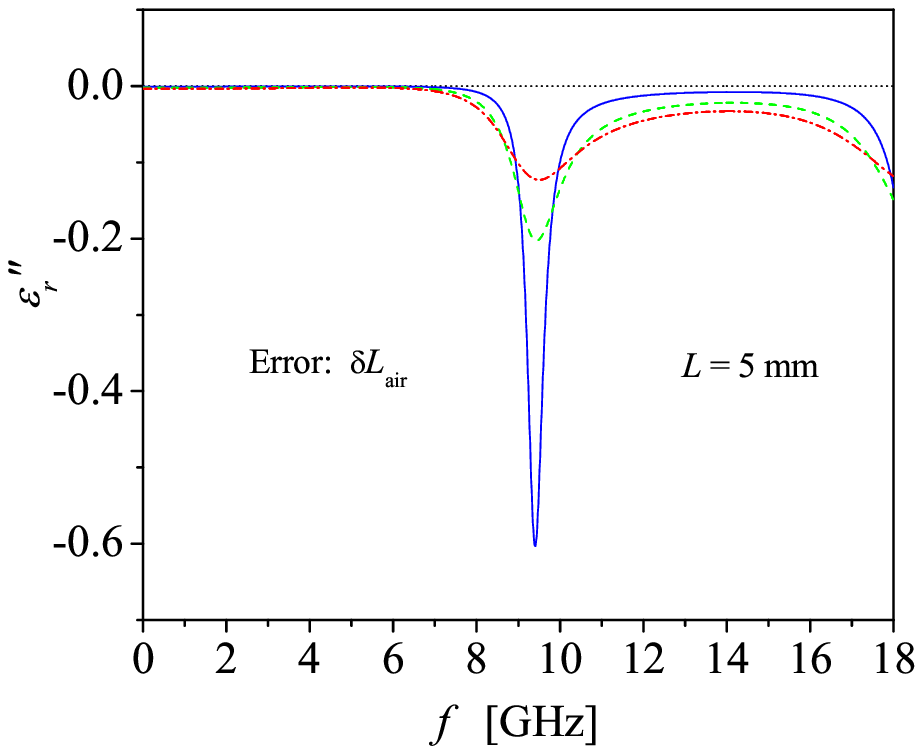}
\caption{Simulated measurements of complex electromagnetic parameters for lossy magnetic materials with $\epsilon_r$ $=$ $5$, $\mu_r$ $=$ $2-0.1i$ (solid-blue), $2-0.3i$ (dashed-green) and $2-0.5i$ (dot-dashed-red). The error due to $L_{air}$ is set to 0.1 mm.}
\label{MagneticHighLoss_u}
\end{figure}

The magnetic energy loss is included by assuming a non-zero value for the imaginary part of magnetic permeability. The simulations are done for three different materials with $\mu_r$ $=$ $2-0.1i$,  $2-0.3i$ and $2-0.5i$. The dielectric loss is taken zero with $\epsilon_r$ $=$ $5 + 0i$. Fig. \ref{MagneticHighLoss_u} shows how the measured results would look like when only the $L_{air}$ error is considered. These plots were made using both $n$ and $z$, $\epsilon_r = n/z$ and $\mu_r = nz$. In principle, it is possible to extract $\mu_r$ as $\mu_{r} = n^{2}/\epsilon_{r}$, which would lead, much like in the case of dielectric  materials, to lower errors. However, the typical situation in practice is that the experimentalist has no reliable knowledge of either $\epsilon_{r}$ or $\mu_r$ for such materials. Similar to the lossy dielectric material previously simulated, the calculations show that there is no divergence in the permittivity and permeability spectra of lossy magnetic materials.

\section{Conclusions}
This paper shows that the electromagnetic properties of materials, such as the complex refractive index, the complex permittivity and the complex permeability, can be measured using the transmission/reflection method without having to know the positions of the reference planes. The method is explicit ({\it i.e.} non-iterative) and easy to implement. The value of magnetic permeability is not assumed: both magnetic and non-magnetic material can be characterized.

The algorithm has been verified experimentally on two types of low-loss dielectric materials. The results show that at Fabry-P\'erot resonances, discontinuities occur in the real-part spectra of constitutive parameters. However, these discontinuities occur only over a few measured points, therefore in many situations of practical interest the results at these points can be neglected and the true values of material parameters can be found by interpolation. For the spectra of imaginary values, the method leads to results which are relatively stable all over the measurement bandwidth.

The numerical analysis and the experiments show that the method has relatively low uncertainties at higher frequencies. This characteristic is an intrinsic property of the transmission/reflection method because a significant change in the phase of the S-parameter due to the presence of the material requires that the sample length should be above a certain size, compared to the wavelength of the signal.

\section*{Acknowledgment}

The authors would like to thank Dr. Tapani Matala-Aho for stimulating discussions, and Kai Poras for helping in the sample 
preparation process.

\end{document}